\documentclass[aps,superscriptaddress,prb, preprint]{revtex4-1}

\label{pre:packages}
\usepackage{siunitx}
\usepackage[normalem]{ulem}
\usepackage{amsmath,amssymb}
\usepackage[inline]{enumitem}
\usepackage{array}
\usepackage{xspace}
\usepackage{graphicx}
\usepackage{color}
\usepackage{etoolbox} 

\usepackage[colorinlistoftodos]{todonotes}

\usepackage[capitalize]{cleveref}

\makeatletter
\appto{\appendix}{%
  \@ifstar{\def\theequation@prefix{A.}}%
  {}%
}

\makeatother

\label{pre:macros}

\newcommand*{\bmb}{$\mathrm{BaMn_2Bi_2}$\@\xspace}
\newcommand*{\bms}{$\mathrm{BaMn_2Sb_2}$\@\xspace}
\newcommand*{\bma}{$\mathrm{BaMn_2As_2}$\@\xspace}
\newcommand*{\bmp}{$\mathrm{BaMn_2P_2}$\@\xspace}

\newcommand*{\bmpn}{$\mathrm{BaMn_2}Pn\mathrm{_2}$\@\xspace}

\newcommand*{\bcoa}{$\mathrm{BaCo_2As_2}$\@\xspace}
\newcommand*{\bcra}{$\mathrm{BaCr_2As_2}$\@\xspace}
\newcommand*{\bcua}{$\mathrm{BaCu_2As_2}$\@\xspace}
\newcommand*{\bna}{$\mathrm{BaNi_2As_2}$\@\xspace}
\newcommand*{\bza}{$\mathrm{BaZn_2As_2}$\@\xspace}
\newcommand*{\btmpn}{$\mathrm{BaTM}_2Pn_2$\@\xspace}
\renewcommand{\vec}[1]{\mathbf{#1}}

\newcommand{\tmin}{T_{\mathrm{min}}}
\newcommand{\tstar}{T^\ast}

\mathchardef\mhyphen="2D
\newcommand{\dphyb}{d\mhyphen p}
\newcommand{\pt}{\mathrm{PT}}
\newcommand{\vecHab}{\vec{H}_{ab}}
\newcommand{\vecH}{\vec{H}}
\newcommand{\ecurrent}{\vec{j}_{\mathrm{E}}}

\newcommand{\bluesout}{\bgroup\markoverwith{\textcolor{red}{\rule[0.4ex]{4pt}{1pt}}}\ULon}

\DeclareSIUnit[number-unit-product = \,]{\unitcell}{\mathrm{unit\ cell}}
\DeclareSIUnit[number-unit-product = \,]{\hole}{\mathrm{hole}}

\newif\ifprint
\label{pre:ifprint}

\printtrue

\ifprint
\colorlet{hl}{green!50!}
\colorlet{green}{green!50!black!}
\colorlet{blue}{blue!80!black!}

\else
\usepackage[prefix=]{xcolor-solarized}
\colorlet{gray}{base01}
\colorlet{hl}{base02}
\makeatletter
\newcommand{\globalcolor}[1]{%
  \color{#1}\global\let\default@color\current@color
}
\makeatother
\AtBeginDocument{
  \globalcolor{base0}
  \pagecolor{base03}
}
\fi

\begin{document}

\title{
  {Large negative magnetoresistance in $\bf{BaMn_2Bi_2}$ antiferromagnet}
}


\author{Takuma Ogasawara} 
\affiliation{Department of Physics, Graduate School of Science, Tohoku University, 6-3 Aramaki, Aoba, Miyagi, Japan}

\author{Kim-Khuong Huynh}
\email{huynh.kim.khuong.b4@tohoku.ac.jp}
\affiliation{AIMR, Tohoku University, 1-1-2 Katahira, Aoba, Sendai, Miyagi, Japan}


\author{Time Tahara}
\affiliation{Center of Advanced High Magnetic Field Science, Graduate School Science, Osaka University, Machikaneyama, Toyonaka, Osaka, Japan}

\author{Takanori Kida}
\affiliation{Center of Advanced High Magnetic Field Science, Graduate School Science, Osaka University, Machikaneyama, Toyonaka, Osaka, Japan}

\author{Masayuki Hagiwara}
\affiliation{Center of Advanced High Magnetic Field Science, Graduate School Science, Osaka University, Machikaneyama, Toyonaka, Osaka, Japan}

\author{Denis Ar\v{c}on}
\affiliation{Faculty of mathematics and physics, University of Ljubljana, Jadranska c. 19, 1000 Ljubljana, Slovenia}
\affiliation{Jozef Stefan Institute, Jamova c. 39, 1000 Ljubljana, Slovenia}

\author{Motoi Kimata}
\affiliation{Institute for Materials Research, Tohoku University 2-1-1 Katahira, Aoba-ku, Sendai 980-8577, Japan}

\author{Stephane Yu Matsushita}
\affiliation{AIMR, Tohoku University, 1-1-2 Katahira, Aoba, Sendai, Miyagi, Japan}

\author{Kazumasa Nagata}
\affiliation{Department of Physics, Graduate School of Science, Tohoku University, 6-3 Aramaki, Aoba, Miyagi, Japan}

\author{Katsumi Tanigaki}
\email{katsumi.tanigaki.c3@tohoku.ac.jp }
\affiliation{AIMR, Tohoku University, 1-1-2 Katahira, Aoba, Sendai, Miyagi, Japan}
\affiliation{BAQIS, Bld. 3, No.10 Xibeiwang East Rd., Haidian District, Beijing 100193, China}

\date{\today}


\begin{abstract}

  A very large negative magnetoresistance (LNMR) is observed in the insulating regime of the antiferromagnet \bmb{} when a magnetic field is applied perpendicular to the direction of the sublattice magnetization.
  High perpendicular magnetic field eventually suppresses the insulating behavior and allows \bmb{} to re-enter a metallic state.
  This effect is seemingly unrelated to any field induced magnetic phase transition, as measurements of magnetic susceptibility and specific heat did not find any anomaly as a function of magnetic fields at temperatures above $\SI{2}{\kelvin}$.
  The LNMR appears in both current-in-plane and current-out-of-plane settings, and Hall effects suggest that its origin lies in an extreme sensitivity of conduction processes of holelike carriers to the infinitesimal field-induced canting of the sublattice magnetization.
  The LNMR-induced metallic state may thus be associated with the breaking of the antiferromagnetic parity-time symmetry by perpendicular magnetic fields and/or the intricate multi-orbital electronic structure of BaMn$_2$Bi$_2$.

  
\end{abstract}

\maketitle

%
\section{Introduction}
\label{sec:introduction}

Numerous exotic materials adopt the tetragonal ThCr$_2$Si$_2$ (122) crystallographic structure [Fig.~\ref{fig:struct}(a)], and more than 400 compounds have been known \cite{hoffmann1985a} to be classified into this structure.
An extended family related to BaFe$_2$As$_2$ also belongs to this large family.
Since the high temperature superconductivity was found for the BaFe$_2$As$_2$ family of compounds by chemically doping or under high pressure \cite{rotter-2008,alireza-2009}, researchers have systematically studied on almost all possible combinations in the formula of BaTM$_2$Pn$_2$, where TM stands for $3d$-transition metals and Pn pnictogen elements. 
The flexibility of the tetragonal (122) structure allows the TM site to be replaced by many $3d$ elements from Cr to Zn, resulting in a wide variety of materials hosting diverse ground states.
For instance, %
\bcra is a G-type (checkerboard) antiferromagnetic (G-AFM) metal with large holelike Fermi surfaces \cite{singh_itinerant-2009,filsinger_antiferromagnetic-2017};
\bcoa is a paramagnetic metal close to a magnetic quantum critical point \cite{sefat_renormalized-2009,dhaka-2013,van_dynamical-2014,ahilan_nmr-2014}; \bcua is a weakly correlated $sp$ metal\cite{singh_electronic-2009,wu_direct-2015};
\bna is a superconductor with a conventional electron-phonon coupling mechanism \cite{subedi_density-2008,kurita_lowT-2009};
and \bza is potentially a host material for diluted magnetic semiconductors \cite{zhao-2013,man-2015,suzuki_photoemission-2015,gu_diluted-2016}.

A subfamily of \bmpn{} ($Pn$: P, As, Sb, and Bi) compounds can be differentiated from the other \btmpn's for their remarkable ground states with a parity-time symmetry \cite{watanabe_magnetic-2017}, orbital-selective magnetism \cite{craco_2018}, and intriguing transport properties \cite{huynh2019}.
Unlike most of the other TM's, where stable \btmpn{} phases are only available with P and As, Mn forms a complete set of \bmpn{} compounds with all pnictogen elements from P to Bi.
Except for the highly insulating \bmp{}, the other \bmpn's, \bma, \bms, and \bmb, exhibit the behaviors of semiconducting-like antiferromagnets with a common G-type antiferromagnetic (AFM) order with high N\'eel temperatures ($T_{\mathrm{N}}$'s) \cite{huynh2019}.
As we reported recently, these compounds interestingly exhibit an unusual and very large magnetoresistance (LNMR) \cite{huynh2019} under a modest magnetic field of $\SI{16}{\tesla}$.
The properties of the LNMR found in the \bmpn{}'s antiferromagnets do not fit to any known models for magnetoresistance, and its mechanism remains to be an open question.
Recent theoretical and experimental studies have pointed out a rather intricate magnetic-electronic nature of these materials \cite{watanabe_magnetic-2017,craco_2018,huynh2019}.
The symmetry of the AFM order of \bmpn{} (point group $4^\prime/m^\prime m m^\prime$) breaks both time reversal (TR) and space inversion (SI) symmetries, but the combined parity-time (PT) symmetry is preserved \cite{watanabe_magnetic-2017}.
The $\dphyb$ states resulting from this magnetic-electronic entangled states reside at the top of the valence band \cite{watanabe_magnetic-2017}.
Therefore the transport properties of the holelike carriers can be highly susceptible to the perturbations of the magnetic order caused by an external magnetic field $\vecH$.
The origin of the LNMR observed for \bmpn's may thus be associated with the breaking of the PT symmetry in the applied $\vecH$ \cite{huynh2019}.
On the other hand, another theoretical study based on dynamical density functional calculations suggests an intricate orbital-selective magnetism \cite{craco_2018}.
Whereas the out-of-plane $d$ orbitals of Mn form strongly Mott localized band residing far from the Fermi level ($E_{\mathrm{F}}$) in an agreement with the G-type AFM order, 
the bands originated from the in-plane orbitals are nonmagnetic.
The $xy$ band is even more interesting because it is not fully insulating even under high Coulomb interaction $U$, resulting in a small density of holelike states at $E_{\mathrm{F}}$.
These states are of the most importance from the viewpoint of transport properties and can be directly related to the LNMR.

Among pnictogen elements, bismuth has the strongest spin-orbit coupling (SOC) and the largest radii of atomic wavefunctions; the latter gives the strongest $\dphyb$ hybridization.
In our previous research, \bmb exhibits the largest LNMR in the smallest range of magnetic field strength ($H$) in relative to other \bmpn{}'s.
Therefore, \bmb could be the key compound for understanding the observed LNMR.
In the present paper, we report the detailed measurements of the electrical transport, thermodynamic, and magnetic properties of \bmb{}.
The isobaric specific heat $C_p$ and magnetic susceptibility $\chi$ shows no detectable structural and/or magnetic transition below $T_{\mathrm{N}}$ even under high $H$. 
Our measurements of the magnetotransport properties include both the in-plane (electric current $\vec{j}$ parallel to the $ab$ conduction layer) and the out-of-plane ($\vec{j}$ along to the $c$-axis) for various directions of $\vecH$ and at different temperatures.
We found that the LNMR manifests in both current-in-plane and current-out-of-plane configurations.
Regardless of the direction of the electric current $\vec{j}$ in relative to the crystal axes, the conductivity of \bmb{} in both current-in-plane and current-out-of-plane settings is maximized when $\vecH$ lies in the $ab$ plane, i.e. perpendicular to the direction of the sublattice magnetization of the G-type AFM order.
The electrical transport of \bmb{} is thus sensitive even to the small $\vecH$-induced in-plane magnetization, the later of which breaks the PT-symmetry.

\section{Results}
\label{sec:results}

\subsection{Electronic structure}
\label{sec:calculation}
\begin{figure}
  \begin{center}
    \includegraphics[scale = .9]{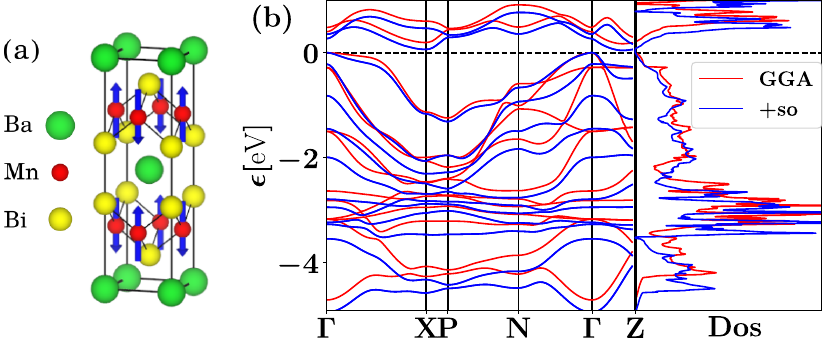} 
  \end{center}
  \caption{(a) crystal structure and G-type antiferromagnetic order of \bmb. (b) band structure and density of states of \bmb with and without spin-orbit coupling.}
  \label{fig:struct}
\end{figure}

In order to understand the electronic structure of \bmb, we performed Density Functional Theory (DFT) calculations employing the WIEN2k code, which implements a full-potential linearized augmented plane-wave LAPW + local orbitals method \cite{wien2k}.
We used the experimental information of the crystal lattice and the magnetic ordering \cite{saparov-2013} as the input parameters (see Sec.~\ref{sec:method}).
To confirm the effect of SOC on the electronic structure, we compare the results of DFT (using GGA approximation) with and without SOC [Fig.\ref{fig:struct}(b)].
Turning on the SOC does not change the electronic structure qualitatively, however the bandgap decreases and a degeneracy of the second hole band is lifted.
Similar calculations for \bma yielded an electronic structure with the effect of SOC being much smaller in line with a smaller SOC of As element, which is in agreement with the experimental results of ARPES measurements \cite{zhang2016}.
We thus conclude that SOC has the largest effect on the electronic structure for \bmb{} within the \bmpn{} family of compounds.

\subsection{Large negative magnetoresistance (LNMR)}
\label{sec:isothermal-lmr}
\begin{figure*}[htbp]
  \begin{center}
          \includegraphics[scale = .9]{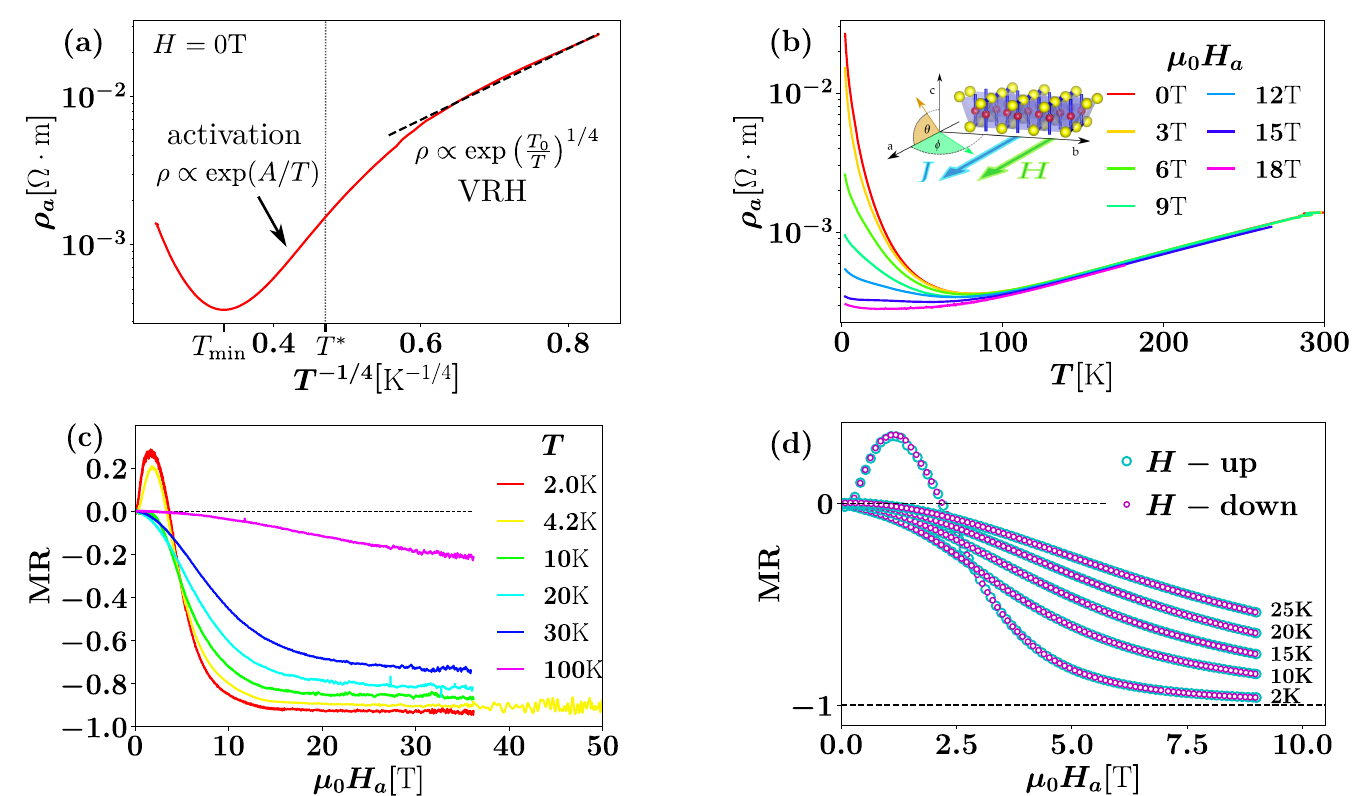}
  \caption{
    (Color online) 
    (a) Temperature dependence of $\rho_{a}$ at $\SI{0}{\tesla}$ plotted as a function of $T^{-1/4}$.
    The dashed line show the fitting using a variable range hopping rule for the data at low temperatures.
    (b) Temperature dependencies of the in-plane electrical resistivity ($\rho_{a}$) of \bmb measured under different strengths of perpendicular magnetic fields ($H_{a}$).
    The electric current $\vec{j}$ was applied along the $a$-axis [inset].
    (c) Longitudinal MR ($\theta=\SI{0}{\degree}$) observed at various $T$'s. 
    (d) No hysteresis was observed in scanning $H_{ab}$ both in increasing and decreasing modes.
  }
  \label{fig:rhoxx-mr}
  \end{center}
\end{figure*}

\subsubsection{Current-in-plane magnetoresistance}
\label{sec:intra-mr}
Fig.~\ref{fig:rhoxx-mr}(a) and (b) show the $T$-dependencies of $\rho_a$ of a BaMn$_2$Bi$_2$ single crystal measured different strengths of $\vec{H}$ parallel to the $a$-axis of the crystal. 
These measurements were done in the in-plane configuration, and both $\vec{j}$ and $\vecH$ were directed along the crystallographic $a$-axis, i.e., perpendicular to the easy $c$-axis of the G-type AFM order.
The dependence of resistivity on temperature of \bmb{} is similar to those found for all \bmpn{}'s, however \bmb{} is much more conductive \cite{huynh2019}.
As shown in Fig.~\ref{fig:rhoxx-mr}(a), the zero-field $\rho_a(T)$ curve first decreases with decreasing temperatures, but then reaches a broad minimum before it starts to increase on further cooling.
This metal-to-insulator cross-over (MIC) occurs at $T_{\mathrm{min}} \approx \SI{83}{\kelvin}$, being apart from the N\'eel temperature $T_\mathrm{N} = \SI{387}{\kelvin}$ \cite{calder-2014}.
Below the MIC, the $\rho_a(T)$ curve can be fitted by an activation law $\rho_a(T) \propto \exp(A/T)$ down to $T^\ast \approx \SI{20}{\kelvin}$ (the fit is not shown).
  At lower temperatures, the $\rho_a(T)$ curve exhibits the rule of Mott's three-dimensional variable range hopping (VRH) described as $\rho \propto \exp \left[(T/T_0)^{1/4}\right]$.
  That suggests the insulating behaviors at $T < \tmin$ is not caused by a bandgap but rather due to a localization of charge carriers.
Interestingly, when $\vec{H} \parallel a$ is turned on, a very large negative MR appears at $T  < \SI{83}{\kelvin}$ and reduces $\rho_a$  significantly.
As a result, the low-temperature insulating trend of $\rho_a$ is suppressed by strong $H_a$, and \bmb{} exhibits a bad metallic behavior at $\mu_0H_a = \SI{18}{\tesla}$.

As shown in Fig.\ref{fig:rhoxx-mr}(c), the LNMR is strongly dependent on temperatures.
At the lowest $T$ achieved in the present measurements, LNMR exhibits a pronounced bell-shaped field dependence.
For instance, at $T=\SI{2}{\kelvin}$, with turning on $H_a$ the longitudinal MR at first increases to a maximum $\mathrm{MR} \approx \SI[retain-explicit-plus]{+40}{\percent}$ at $H_a \approx \SI{1.7}{\tesla}$.
As $H_a$ increases, MR drops quickly to the negative values ending up by bending to a saturation of $\mathrm{MR} \approx \SI{-96}{\percent}$ at $\mu_0H_a \approx \SI{16}{\tesla}$.
This corresponds to the metallic state being re-entered in the $\rho_a(T)$ curve.
With further increase of $H_a$, an almost constant MR can be realized up to the highest $H_a$ of $\SI{55}{\tesla}$.
As $T$ increases, the positive component in MR fades away and disappears at $T^{\ast} \approx \SI{20}{\kelvin}$, leaving only the negative MR at elevated $T$'s.
The values of the saturated negative MR also reduce with increasing $T$ and stronger $H_a$'s are needed for returning the MR back to the same saturation realized at lower $T$.
In Fig.\ref{fig:rhoxx-mr}(d), the MR curves under increasing and decreasing $H_a$'s totally overlap on each other, and no hysteresis was observed.
This suggests that the origin of the LNMR is apart from a $\vecH$-induced first order phase transition \cite{tian2015}.

\begin{figure*}
  \begin{center}
    \includegraphics[scale = .9]{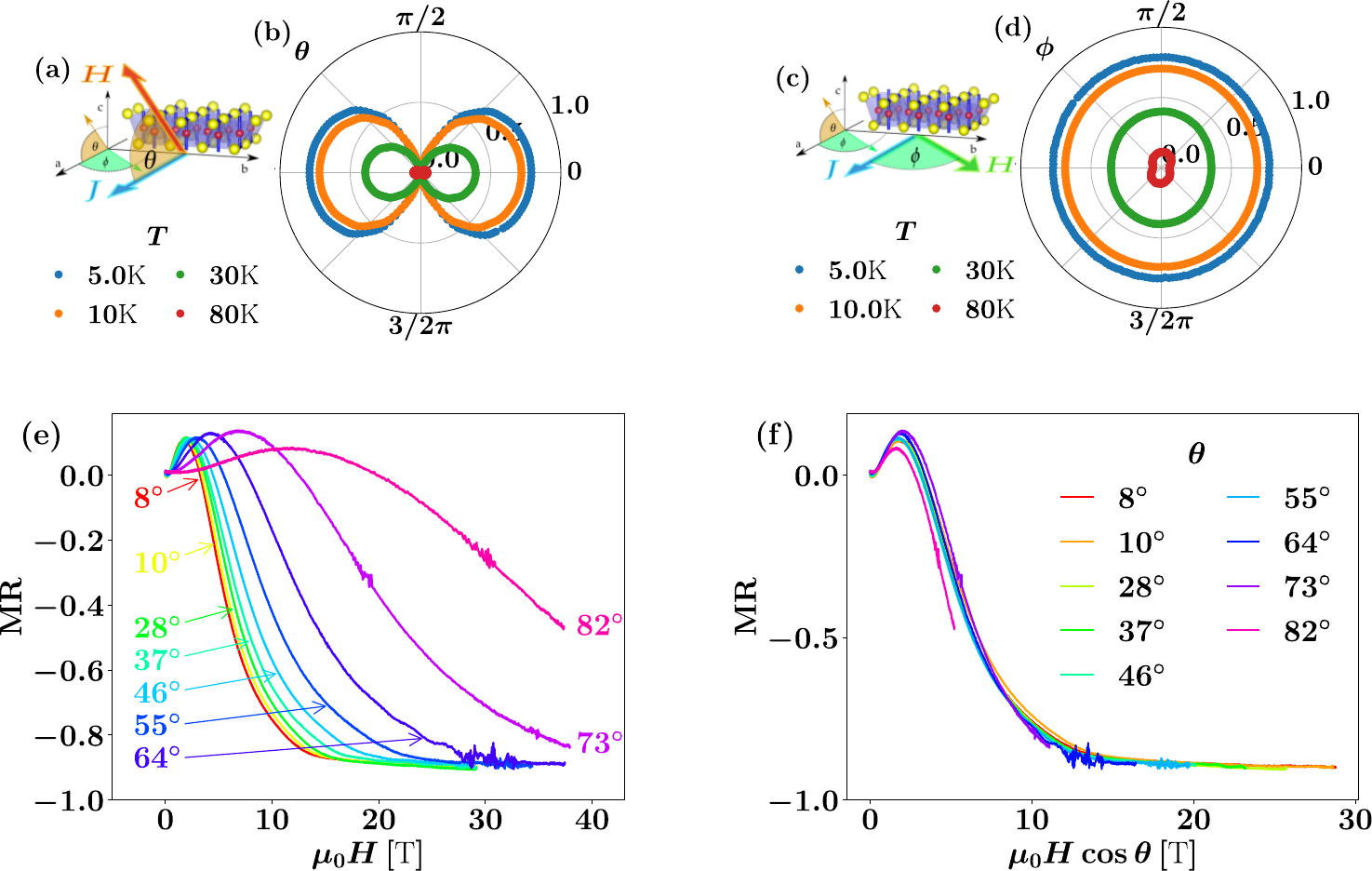}
  \end{center}
   \caption{
   (a) The setting for the measurements of angle resolved MR with current-in-plane in $\theta$ rotation of $H$. 
   (b) The absolute values of MR as function of $\theta$ in polar plot. 
   (c) and (d) are same of (a) and (b) in $\phi$ rotation of $H$.
   (e) The magnetic field dependence of MR for various $\theta$s. 
   (f) MR scaled by in-plane magnetic field $\mu_0 H \cos\theta$}
  \label{fig:intra-theta-phi}
\end{figure*}

The LNMR in \bmb shows an interesting angular dependence on the direction of $\vecH$.
Figs. \ref{fig:intra-theta-phi}(a) and (b) show that the LNMR is very sensitive to the angle $\theta$ that $\vec{H}$ makes with the $ab$-plane.
A $\vec{H}$ aligned within the $ab$ plane maximizes the LNMR, and with $\vec{H}$ rotating away from basal plane, the magnitude of the LNMR quickly reduces.
Individual $H_a$ scans at different $\theta$ angles displayed in Fig. \ref{fig:intra-theta-phi}(e) show that the MR curves are elongated toward higher $H_a$ values as $\theta$ increases.
This strong out-of-plane anisotropy results in a curious horizontal eight-figure shape as the absolute value of the LNMR is plotted against $\theta$ in the polar coordinate representation shown in Fig. \ref{fig:intra-theta-phi}(a).
On the contrary, the LNMR effectively remains in its maximized value as $\vec{H}$ rotates within the $ab$-plane [Figs. \ref{fig:intra-theta-phi}(e)].
Changing the $\phi$-angle from $\SI{0}{\degree}$ to $\SI{90}{\degree}$, i.e., from being parallel to anti-parallel with the current $\vec{j}$ and the $a$ axis, yields only an insignificant angular dependence.
The angle resolved measurements suggest that the component of the magnetic field lying in the crystallographic $ab$-plane and perpendicular to the AFM sublattice magnetization, $\vec{H}_{ab}$, plays a crucial role in the mechanism of the LNMR.
In Fig.~\ref{fig:intra-theta-phi}(f), when the MR curves collected at different $\theta$ angles are plotted versus the in-plane component of magnetic fields, i.e. $H_a = H\cos\theta$, almost all the MR curves scales to a single $H_a$-dependence.
However, there are some discrepancies coming from the MR curves measured at large $\theta$'s, at which the out-of-plane component $H_c$ of the magnetic field is larger than the in-plane $H_a$.
These may indicate that the positive component of the MR has an additional dependence on $H_c$.
This is different from the $H$-dependence of the negative MR component, which shows a very good scaling with respect to $H\cos\theta$.

\subsubsection{Current-out-of-plane magnetoresistance}


\label{sec:inter-mr}

\begin{figure}
  \begin{center}
  \includegraphics[scale = .9]{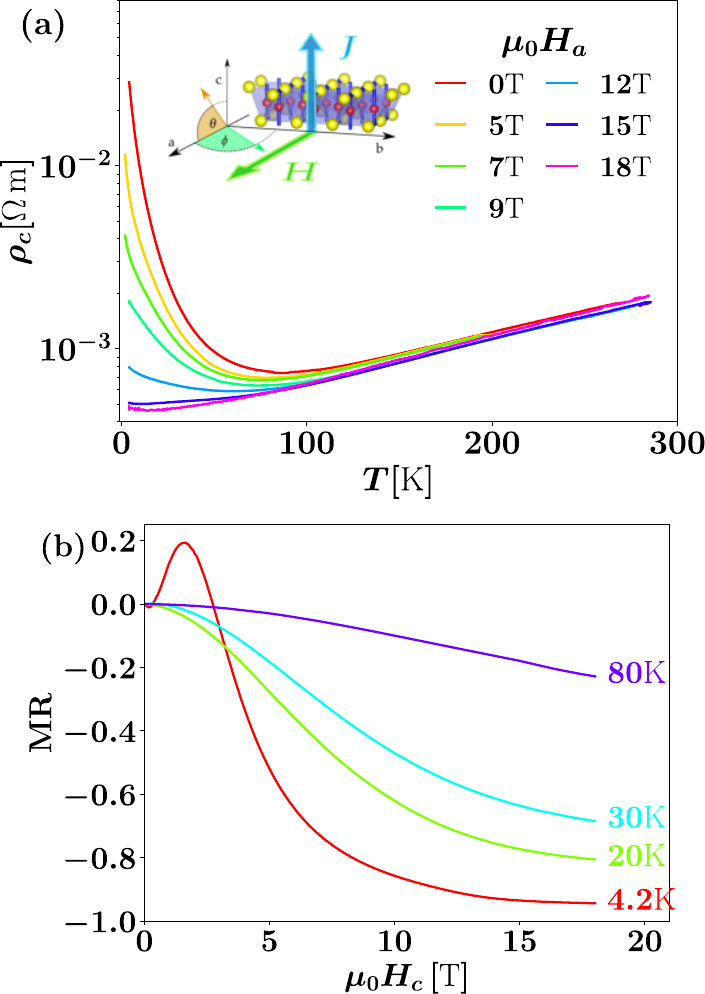}
\end{center}
  \caption{
    (Color online)
    (a) Temperature $(T)$ and magnetic field $(H)$ dependence of electrical resistivity ($\rho_{c}$) of \bmb{}  when the electric current $\vec{j}$ is parallel to the $\hat{c}$-axis (out-of-plane geometry).
    (b) MR observed at various $T$'s.
    Negative MR becomes larger with a decrement in $T$. 
    The positive MR is only observed at $T$s lower than $T^{*}$ under weak $\mu_0H$.
  }
  \label{fig:rhoc-mr}
\end{figure}

\begin{figure*}
  \begin{center}
    \includegraphics[scale = 0.9]{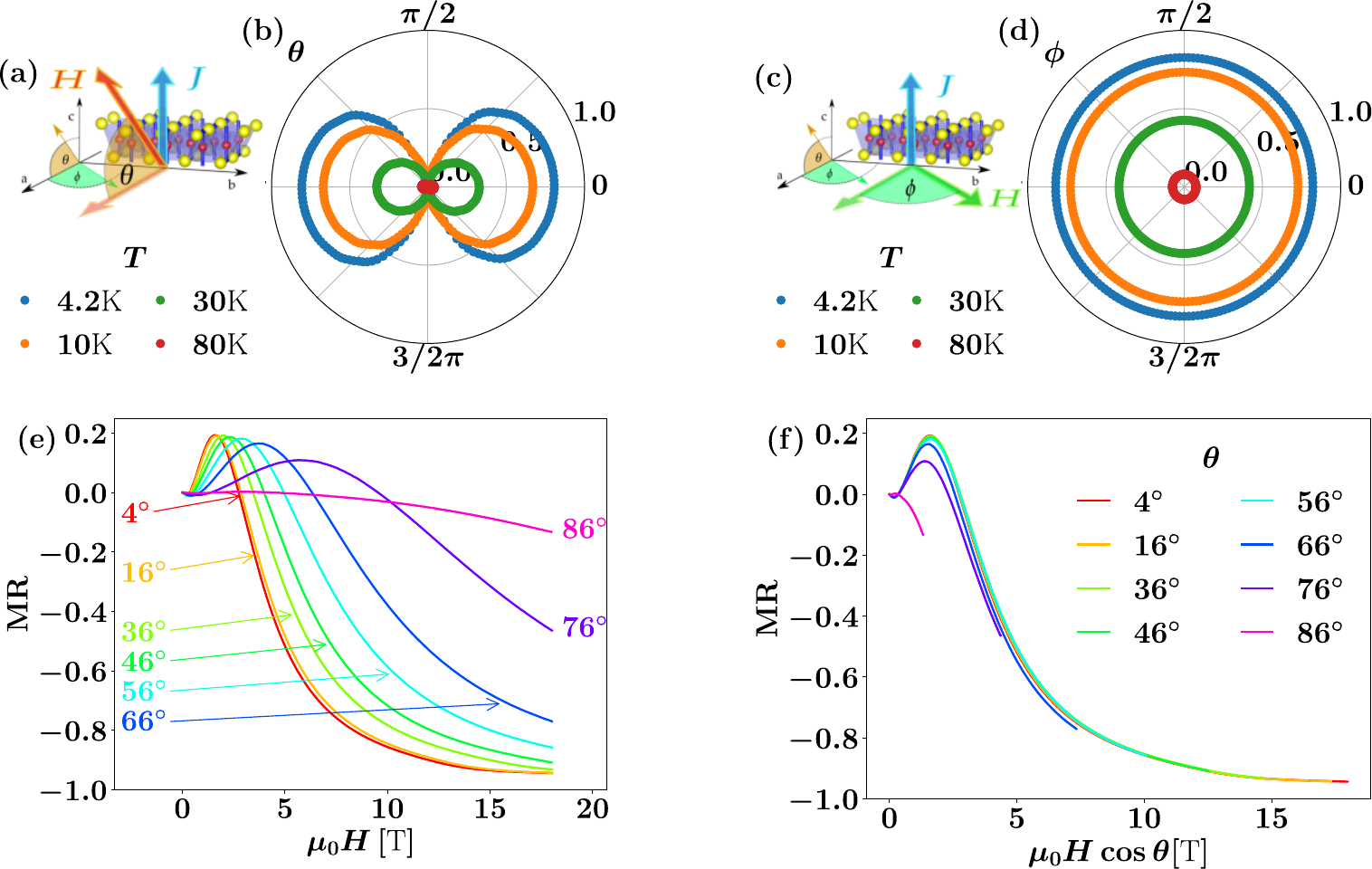}
  \end{center}
 \caption{
   (a) The setting for the measurements of angle resolved MR with current-out-of-plane in $\theta$ rotation of $H$. (b) The absolute values of MR as function of $\theta$ in polar plot. 
   (c) and (d) are same of (a) and (b) in $\phi$ rotation of $H$.
   (e) The magnetic field dependencies of MR for various $\theta$s.
   (f) MR scaled by in-plane magnetic field $\mu_0 H \cos\theta$.}
  \label{fig:inter-theta-phi}
\end{figure*}

The angular dependence of the LNMR described in the previous section suggests that the magnetoconductivity of \bmb{} is mainly determined by the strength of the in-plane magnetic field $\vec{H}_{ab}$, being regardless of the angle between $\vec{H}_{ab}$ and the electric current $\vec{j}$.
In order to confirm this, we carried out current-out-of-plane measurements with $\vec{j}$ flowing along the $c$-axis, i.e. being parallel to the easy axis of the AFM.
Figs.~\ref{fig:rhoc-mr}(a) and (b) show that the LNMR also appears under the $\vecH_a$ with a dependence on the field strength $H_a$ and temperatures being very similar to that seen in the in-plane LNMR ($\vec{j} \parallel ab$).
Fig.~\ref{fig:inter-theta-phi} shows additional experiments for angular-dependence of the out-of-plane LNMR under rotating $\vecH$.
The results follow the same trend with those observed in current-in-plane settings, i.e. the LNMR is very anisotropic with respect to the $\theta$ angle and show only a small modulation versus $\phi$.

As described so far, the LNMR as a function of $T$ and $\vec{H}$ is very similar between current-in-plane and current-out-of-plane settings.
It is solely determined by the direction of $\vecH$, and achieves the largest value when $\vecH$ is perpendicular to the sublattice magnetization of the AFM ground states with the PT symmetry.
These experimental results may imply that the LNMR is likely associated with the breaking of symmetry in the AFM ordered ground state of this system.


\subsection{Transverse Hall effects}
\label{sec:transv-hall-effects}

\begin{figure}
  \begin{center}
  \includegraphics[scale = 0.9]{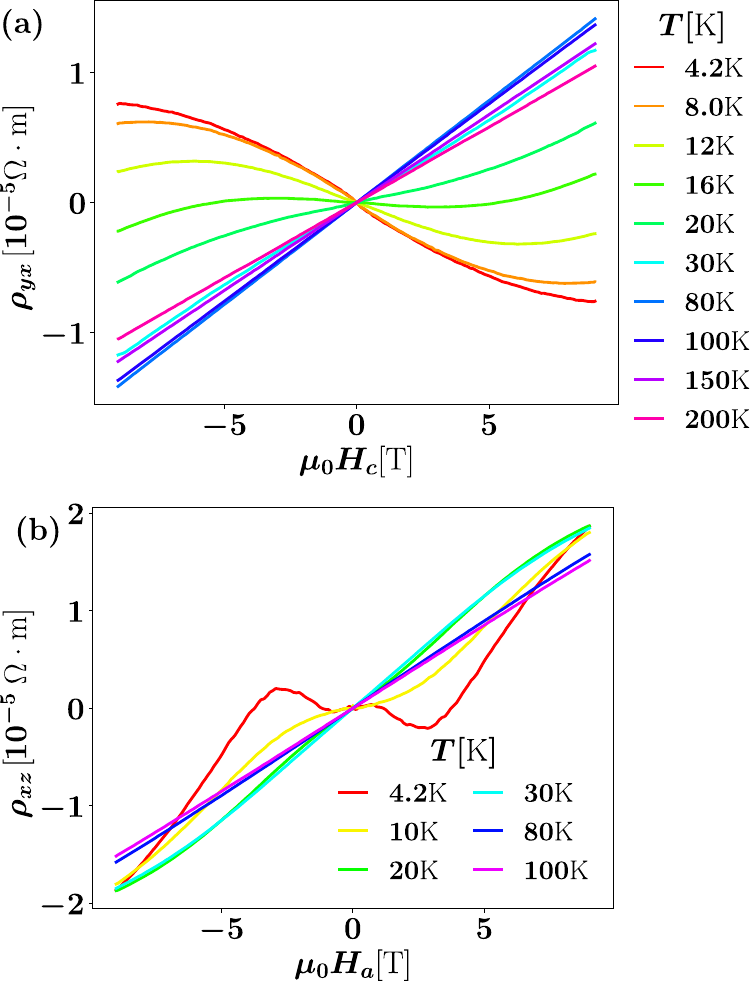}
\end{center}
  \caption{
    (Color online) $H_c$-dependences of (a) $\rho_{yx}$ (current-in-plane)  and (b) $\rho_{xz}$ (current-out-of-plane)  measured at various $T$'s under $\vec{H} \parallel \hat{c}$ and $\hat{a}$, respectively.}
  \label{fig:hall}
\end{figure}

The transverse Hall resistivity for both in-plane and out-of-plane configurations was measured to investigate the temperature dependence of carrier characteristics.
In the in-plane (out-of-plane) configuration, $\rho_{yx}$ ($\rho_{xz}$) was measured as $\vecH$ was parallel to the $c$-axis ($a$-axis) and $\vec{j}$ was along the $a$-axis ($c$-axis) at different $T$'s.
The in-plane configuration yields an $\rho_{yx}$ that increases linearly with $H_c$ for $T \geq \SI{20}{\kelvin}$, indicating that a single holelike carrier type is dominant [Fig.~\ref{fig:hall}(a)].
The linear $H_c$-dependence, however, gradually turns into a non-linear one as $T$ decreases further.
The slopes of $\rho_{yx}(H_c)$ curves around $H_c \approx 0$ decreases to zero and eventually crosses to the negative side at $T^{\ast} \approx \SI{20}{\kelvin}$,  while they remain bending towards the positive side under high $H_c$.
The onset of the non-linear $\rho_{yx}(H_c)$ occurs at $T^\ast$, where the positive component in MR simultaneously starts to appear.

\begin{figure}
  \begin{center}
    \includegraphics[scale = 0.3]{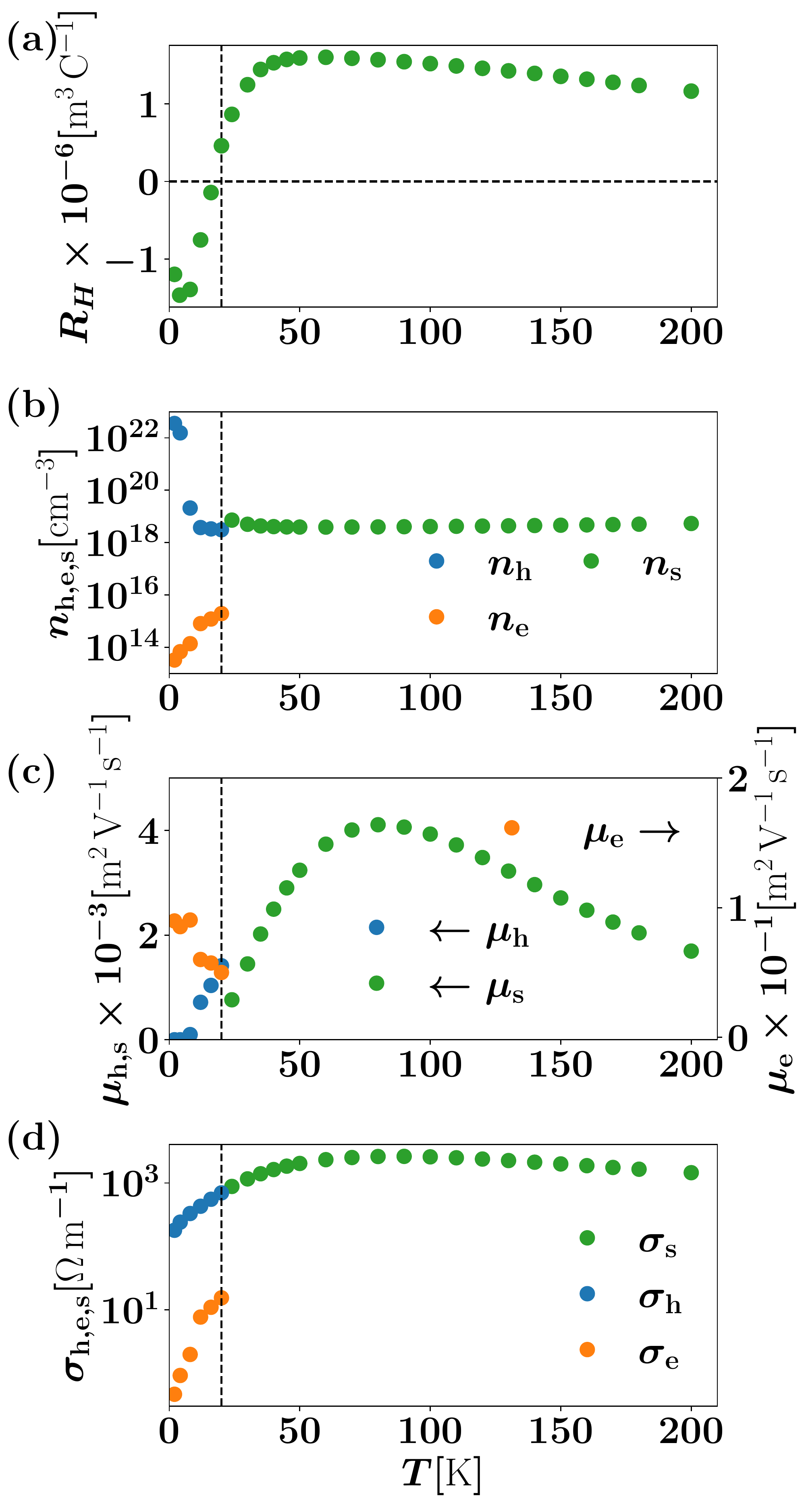}
\end{center}
  \caption{(Color online) Temperature dependence of (a) Hall coefficient calculated from the slope near the zero-field region of Hall resistance. (b) Hall carrier density, (c) mobility and (d) effective conductivity calculated from the fitting of Hall resistivity utilizing a single carrier model at $T>\SI{20}{K}$ and a two-carrier mode at $T\leq \SI{20}{K}$, respectively. 
  We note that the reliability of the data points at $T\leq \SI{20}{K}$ can be questionable due to the validity of the two-carrier-type model in the context of BaMn$_2$Bi$_2$ (see text).}
  \label{fig:carrier}
\end{figure}

In order to estimate the mobility and number of the carriers at $T > T^{\ast}$, we fitted the $\rho_{yx}(H_c)$ curves using a linear one-carrier-type model, respectively.
In Fig. \ref{fig:carrier}(b) the fitting results show that the holelike carrier number $n_h$ almost does not change with $T$ even across the MIC at $\tmin$. 
On the other hand, the MIC seemingly corresponds to a broad peak in the holelike mobility $\mu_h$ [Fig.~\ref{fig:carrier}(c)], at which its behavior changes from increasing to decreasing with cooling down.
For the data at $T < T^{\ast}$, since a non-linear Hall effect is often interpreted as a signature of a competition between electronlike and holelike carriers \cite{huynh2014}, we tried a semiclassical two-carrier-type fitting [see Sec.\ref{sec:method}].
The results of the two-carrier-types fits suggest an abrupt emergence of electronlike carriers with much lower number ($n_e$) and higher mobility ($\mu_e$), and an increase of $n_h$ as $T$ decreases.
Such changes in carrier numbers are usually accompanied by a magnetic phase transition or structural distortion, but none of which can be detected (see sections \ref{sec:therm-magn} and \ref{sec:magnetism}).
Furthermore, attempts to fit the data using the two-carrier-type model in the temperature window between $\SI{20}{\kelvin}$ and $\SI{40}{\kelvin}$ yield scattered values of number and mobility for the electronlike carriers.
  This is due to the small curvature of the $\rho_{yx}(H_c)$ in this temperature window and the highly nonlinear formula for the two-carrier-type Hall effect shown in Eq.~(\ref{eq:twocarrier}).
The two-band analysis, and the emergence of the electronlike carriers as its result, may not be a reliable estimation for the current case.
The nonlinear Hall effect may come from another reason outside the framework of the semiclassical theory.

As shown in Fig.~\ref{fig:hall}(b), the current-out-of-plane Hall effect is similar to that obtained from the current-in-plane settings for $T > T^\ast$, the positive slopes of the linear $\rho_{yz}$ curves consistently indicate the dominating contribution of holelike carriers.
However, at $T < T^{\ast}$, the current-out-of-plane setting yields more complex Hall resistivity curve.
Here $\vec{H} \parallel a$ and $\vec{j} \parallel c$, and the complex 2-component LNMR simultaneously manifest in this configuration.
Therefore $\rho_{xz}(H_a)$ can be sensitively influenced by the strength of the $H_a$.
On the other hand, the $\rho_{xz}(H_a)$ curves in this regime approach to a linear line with positive slope under high $H_a$'s.
In the $T > T^{\ast}$ regime, the Hall effect is positive and linear, indicating that holelike carriers govern the transport properties.

\subsection{Thermodynamics}
\label{sec:therm-magn}
\begin{figure}[h]
  \begin{center}
  \includegraphics[scale = 0.9]{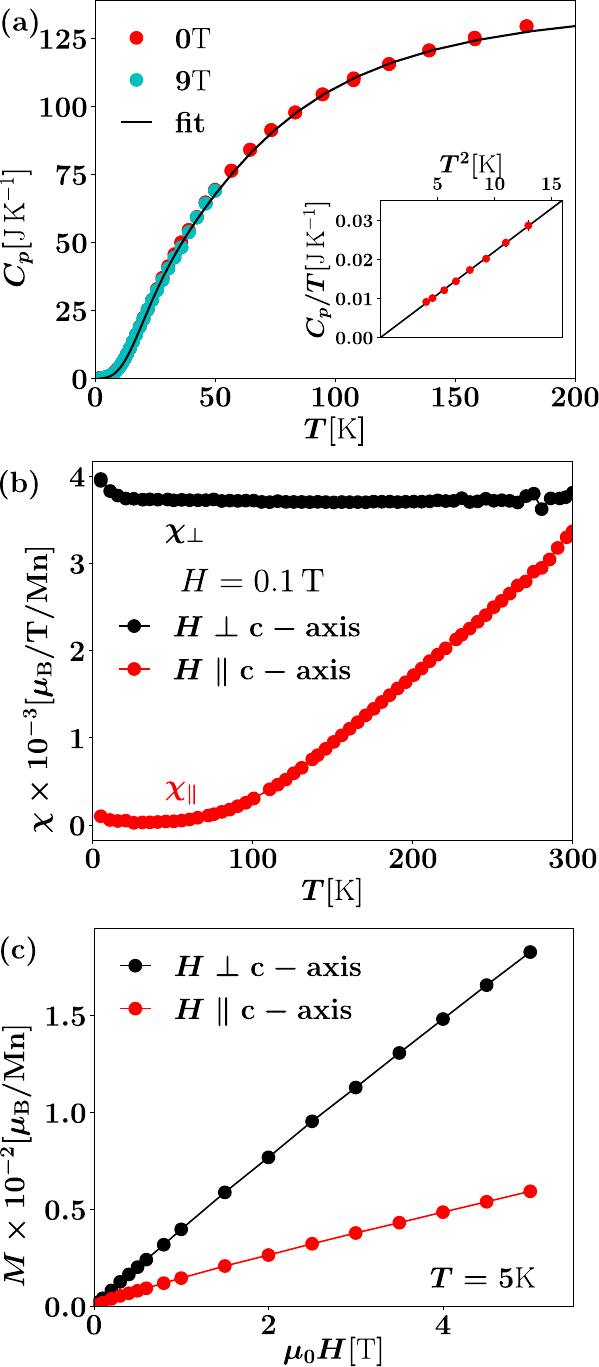}
\end{center}
  \caption{
    (Color online)
    (a) Temperature dependence of specific heat ($C_{P}$) of BMBi. 
    Analyses were made by employing the Einstein and the Debye model.
    Inset is the $C_p/T$ v.s. $T^2$ plot (red circle) and its theoretical fitting (black line) using the equation of $\gamma T + \beta T^3$.
    (b) Anisotropic temperature dependence of magnetic susceptibility ($\chi$) evaluated at $\mu_0H \, = \, \SI{0.1}{\tesla}$ applied along the $\hat{a}$-axis and the $\hat{c}$-axis.
    (c) Magnetic field dependence of magnetic moment ($M$) per Mn ion.}
  \label{fig:thermodynamics}
\end{figure}

In order to investigate other additional effects of the in-plane magnetic fields to the electronic properties, we measured the isobaric specific heat $C_p$ under $\mu_0H_a = \SI{0}{\tesla}$ and $\SI{9}{\tesla}$ [Fig.~\ref{fig:thermodynamics} (a)].
We observed no signature of phase transition and the $C_p(T)$ curve is well-described by a summation of the Debye and the Einstein terms, $C_{p} \approx C_v^{\mathrm{D}} +  C_v^{\mathrm{E}}$ (see sec.~\ref{sec:method}).
A linear fit to $C_p/T$ plotted versus $T^2$ at $T \leq \SI{5}{\kelvin}$ yields an almost zero Sommerfeld coefficient $\gamma \approx \SI{0}{\milli\joule\per\kelvin\squared}$ [Fig.~\ref{fig:thermodynamics}(a), inset].
This indicates a very small number of free carriers in this system, and it is reasonable considering the small number of charge carriers found by transport properties.


\subsection{Magnetism}
\label{sec:magnetism} 
In Fig.\ref{fig:thermodynamics}(b) we show the $T$-dependence of the magnetic susceptibilities $\chi_{\perp}(T)$ and $\chi_{\parallel}(T)$ measured under $\vec{H}$ perpendicular and parallel to the easy $c$-axis, respectively.
The difference between $\chi_{\perp}$ and $\chi_{\parallel}$ shows a typical anisotropy of a colinear AFM between the $ab$-plane and the $\hat{c}$-axis (easy axis).
The N\'{e}el temperature $T_\mathrm{N}$ estimated from the extrapolation is comparable to that obtained from previous studies \cite{saparov-2013,calder-2014}.
We note that a Curie tail was observed in both $\chi_{\perp}(T)$ and $\chi_{\parallel}(T)$.
We found that the magnitude of this component is different among samples, and the older samples show the larger Currie tails.
This reminds us about the mild sensitivity of \bmb{} with respect to the moisture \cite{saparov-2013}.
At the moment, we thus attribute the Curie tail to the contribution of the degradation of the samples.
On the other hand, the $H$-dependence of magnetization ($M$) at $T = \SI{5}{\kelvin}$ is linear up to $H = \SI{5}{\tesla}$ [Fig.\ref{fig:thermodynamics}(c)].
No indication of phase transition is also given from the viewpoint of magnetization measurements as well as specific heat measurements.

\section{Discussion}
\label{sec:discussion}
The most important feature in the transport properties of \bmb is MIC seen in the $T$-dependence of the resistivities (Figs.~\ref{fig:rhoxx-mr} and Figs.~\ref{fig:rhoc-mr}) at $\tmin \approx \SI{83}{\kelvin}$ being far from the N\'eel temperature $T_{\mathrm{N}}$.
Our thermodynamic and magnetic measurements did not reveal any signatures of a structural or magnetic transition even at high $H_{ab}$.
On the other hand, a recent theoretical calculation shows that the MICs observed in \bmpn{} antiferromagnets are the consequences of a Mott localization of holelike states around the $E_{\mathrm{F}}$ \cite{craco_2018}.
At low temperatures, the strong Mott localization results in the insulating behaviors observed in the transport resistivities.
Its strength is weakened by increasing temperature, so that a cross-over to metallic behaviors is expected.
For the case of \bmb{} shown here, the dependencies on temperature of the carrier number and the mobility (Fig.~\ref{fig:carrier}) suggest that the MIC is caused by a vanishing mobility $\mu_h$ of holelike carriers that occupy the states at the top of the $\dphyb$ valence band.
This is qualitatively in a good agreement with the picture of Mott localized holelike carriers shown in Ref. [\onlinecite{craco_2018}].
Interestingly, this Mott-like MIC can be reversed via the LNMR under $\vecHab$, and a metallic behavior appears in the resistivity curves in the whole $T$-range.
If the carrier number is unaffected by $\vecHab$, then the LNMR corresponds to an enhanced $\mu_h$ at all temperatures below $\tmin$.
The metallic state under high $\vecHab$ is accordingly equivalent to a metallic $T$-dependence of $\mu_h$, i.e. $\mu_h$ increases as $T$ decreases.
We note that in order to confirm this scenario, a direct measurement of both carrier number and mobility as a function of $\vecHab$ is necessary.

A key feature in the transport properties of \bmb{} is the anisotropy of the resistivity tensor for various directions of $\vec{j}$ and $\vecH$.
In both current-in-plane and current-out-of-plane settings, the LNMR is maximized when $\vec{H}$ is  perpendicular to the AFM N\'eel vector, the later is pinned along the $\hat{c}$-axis by a large exchange interaction $H_{\mathrm{exchange}}\approx\SI{250}{\tesla}$.
Under $H_{ab} \approx \SI{16}{\tesla}$, \bmb{} can be regarded as a metal in both $\rho_{a}$ and $\rho_{c}$ thanks to the saturation of the LNMR; nevertheless the AFM sublattices magnetization are merely canted a infinitesimal angle in these magnetic fields.
The extreme sensitivity of the MIC to a minute canting of the G-AFM is intriguing, and its mechanism may involve the multi-orbital nature gaining its importance at the vicinity of a Mott transition of \bmb{}\cite{craco_2018}.
Another appealing mechanism may come from the breaking of AFM protected PT symmetry via $\vecHab$.
In this scenario, even a small canting of the AFM moments breaks $\pt$ symmetry and the metallic behavior under $\vecHab$ represents a state of a broken symmetry.

Finally, we would like to come back to the $T$-dependencies of the carrier numbers and the mobilities obtained from our analyses of the transverse Hall effect under $H_{ab} = 0$.
The number of the dominant  holelike carriers estimated from the Hall coefficients at $T > \tstar$, i.e. in the linear regime of the $\rho_{yx}$ curves, is approximately $\SI{4e19}{\per \cubic \centi \meter}$, or $\num{1.2e-3}$ holes per unit cell.
Given the DOS of the calculated band structure, the estimated density of holes is equivalent to an $E_{\mathrm{F}}$ at about $\SI{10}{\milli \electronvolt}$ lower than the top of the valence band.
Such placement of $E_{\mathrm{F}}$ is consistent with the fact that the holelike carrier number stays almost constant across $\tmin$ and suggests that \bmb can be regarded as a highly degenerated semiconductor.
The reduction of $\mu_h$ with decreasing $T$ attributed as the cause of the MIC is clearly seen in this linear regime of $\rho_{yx}$.
At $T < \tstar$, if we ignore the non-linear $\rho_{yx}$ curves by assuming that the number of holelike carrier is constant, we will obtain a $\mu_h$ being decay to zero with decreasing $T$, i.e. being similar to Fig.~\ref{fig:carrier}(c).
On the other hand, the non-linear $\rho_{yx}$ curves with a negative slope in this regime is rather curious.

  A nonlinear Hall effect is often interpreted as a signature of a  two-carrier-type system.
  The results of our analysis using  a conventional two-carrier-type model suggests an abrupt occurrence of an electron-like carrier type, but the conclusion has not presently been supported by any experimental observations such as magnetic and thermodynamic measurements.
Alternatively, there is another explanation  for a negative Hall effect, which may be more plausible in the case of \bmb.
Fig.~\ref{fig:rhoxx-mr}(a) shows that, at $T \leq T^\ast$, the temperature dependence of the resistivity of \bmb enters a VRH regime, i.e. $\rho \propto \exp\left[(T_0 / T)^{1/4}\right]$.
Being caused by the hopping motions in a close path, even holelike carriers can exhibit a negative Hall effect in this regime \cite{friedman1971, mott2012}.
As the temperature increases, the VRH ceases to be valid allowing the Hall effect to gradually change to back to positive.
In this scenario, the electronlike carrier does not exist, and what we observed in the Hall effect is holelike carrier in the VRH regime.
The positive component of the MR observe in \bmb may also arise from the VRH conduction of holelike carriers under magnetic fields \cite{kurobe1982}.


\section{Summary}
We reported transport, thermodynamic, and magnetic properties of AFM \bmb single crystals.
\bmb exhibits a MIC being far away from the N\'eel temperature which is not related to any structural and/or magnetic transitions, as shown by the measurements of specific heat and magnetic susceptibility.
In the insulating regime, the conductivity of \bmb exhibits a very large enhancement when a magnetic field is applied perpendicular to the N\'eel vector of the AFM order and a metallic state seems to be fully recovered, which is the origin of the observed negative LNMR.
Our analyses of the electrical transport properties showed that the MIC and the LNMR are likely due to the change in carrier mobility varying with $T$ and $H_{ab}$.



\section{Method}
\label{sec:method}
\subsection{DFT calculation}
Wien2k package was used to calculate the electronic structure of \bmb. The lattice parameters $a \, = \, b$, $c$ and $z_{Bi}$ are refered from previous study\cite{saparov-2013}. The generalized gradient approximation (GGA) by Perdew, Burke,
and Ernzerhof (PBE)\cite{ggapbe} exchange-correlation potential was choosen. The radii of the muffin-tin sphere
RMT were 2.5 bohrs for all atoms. A $15 \times 15 \times 15$ k-point mesh was utilized in the self-consistent calculations. The truncation of the modulus of the reciprocal lattice vector $K_{max}$ , which was used for the expansion of the wave functions in the interstitial regions, was set to $RMT \times K_{max} = 7$. All calculation was done with assumption of G-type AFM ordering. DFT calculation including spin-orbit coupling via the option of Wien2k package was also conducted. The other DFT calculation package, Quantum Espresso, was also used for comparison and Any significant difference were not found. 

\subsection{Synthesis}

\begin{figure}[htbp]
  \includegraphics[scale = .45]{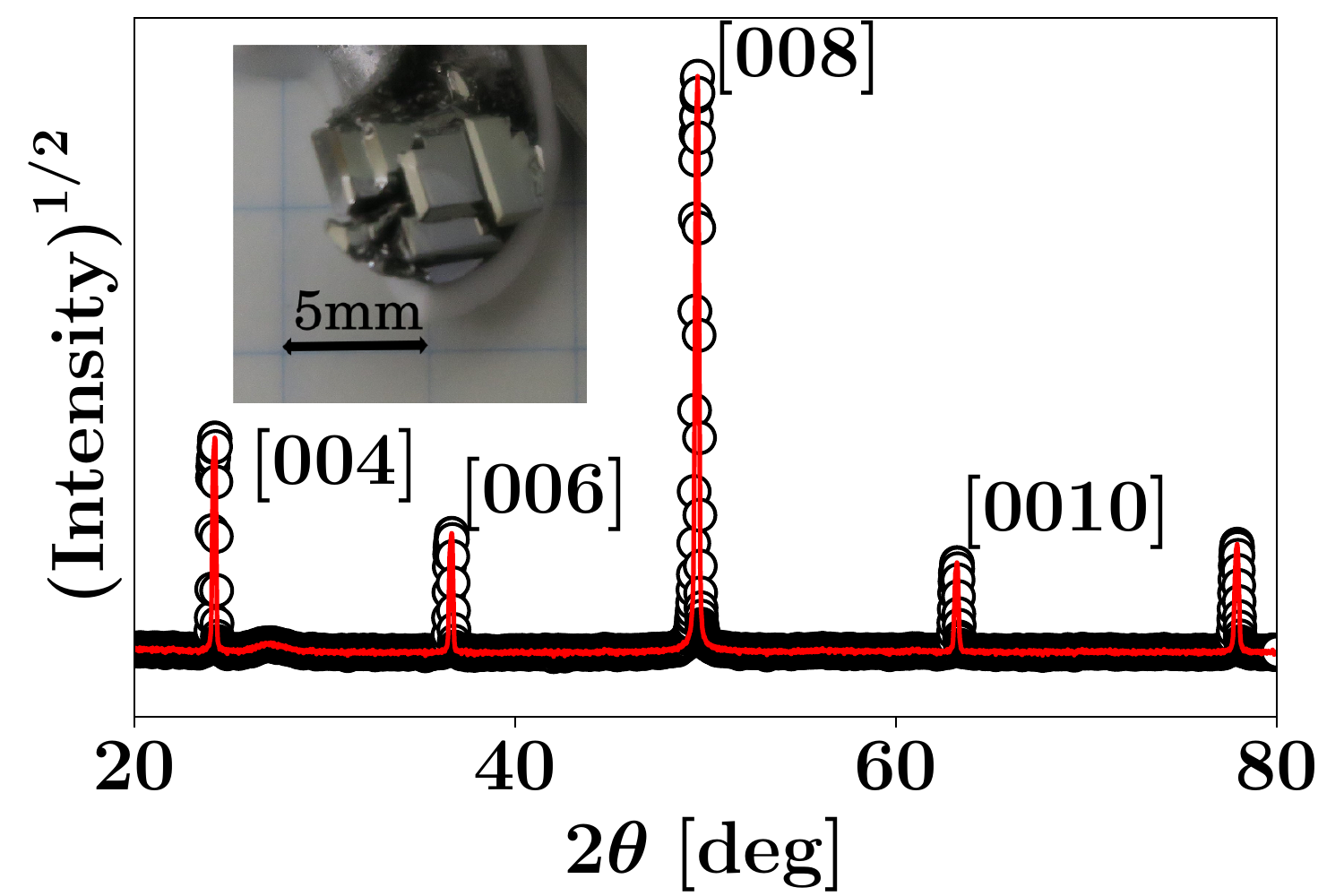}
  \caption{The crystalline X-ray diffraction spectrum along $\hat{c}$-axis. Inset is a photo of single crystals grown in alumina crucible.}
  \label{fig:xrd}
\end{figure}

Single crystals of  \bmb \,  were grown out from the metallic Bi-flux following the method described in Ref.~\onlinecite{saparov-2013}.
In brief, a mixture of Ba, Mn, and Bi with the molar ratio $\mathrm{Ba:Mn:Bi = 1:2:5}$ was loaded into a Al$_2$O$_3$ crucible and then was sealed in a fused-silica tube under helium atmosphere.
The ampule was put into an electric furnace and the temperature was increased to $\SI{1000}{\celsius}$ in $10$ hours, then kept for $10$ hours to ensure complete reactions between the chemicals.
The temperature was then slowly cooled down at the rate of $\SI{2}{\celsius\per\hour}$ to 500 $^\circ$C, at which centrifugal decanting was performed to remove the excess Bi flux.
Large single crystals of centimeter-size and mirror $ab$ surfaces were obtained from the described growth process as shown in inset of Fig.~\ref{fig:xrd}.
It is noted that \bmb \,  single crystals are fairly sensitive to oxygen and moisture in air \cite{saparov-2013} and therefore were kept either in vacuum or in an Argon-filled glovebox.

\subsection{Single Crystal Characterization}
\label{sec:result-xtal}
In order to characterize our single crystals, X-ray diffraction (XRD) measurement done with the help of a RIGAKU SmartLab X-ray Diffractometer. The chemical composition of the single crystals were confirmed by energy dispersive X-ray spectroscopy (EDX) measurement employing a JEOL JSN-7800 scanning electron microscope quipped with an Oxford X-MAX50 analyzer.
Fig.\ref{fig:xrd} shows the result of Crystal XRD and the picture of single crystal. [00l] (l is even due to the body-centered structure) peaks were obtained. The lattice constant is $c = 14.45$ which is comparable to previous work by Saparov $et.al$, $14.687$. The broad peak around 26$^\circ$ is due to the oxidation on the surface. It can be seen in EDX. \par
The result of EDX indicates the inclusion of a little amount of oxygen due to the oxidation on the surface. The ratio after the oxidation subtraction is close to the ideal one i.e. Ba:Mn:Bi = 1:2:2. The fluctuation of ratio is below 1\%.
\subsection{Measurements and analysis}
\label{sec:measurements}

\label{loc:measurements-electrial}
In order to prepare the samples for the measurements, \bmb \,  single crystals were cleaved to remove excessive flux and to expose fresh $ab$-surfaces.
Due to the sensitivity to air of the compound, care were taken so that the samples were exposed less than $2$ hours in the air.
The crystals selected for transport measurements were cut into rectangular shapes whose edges were aligned along the $a$ and $b$ .
For the measurements of longitudinal resistivity ($\rho$) and Hall resistivity ($\rho_{yx}$), six silver-paste electrodes were made on the sample so that the electric current $\ecurrent$ could be applied along the $\hat{a}$-axis.
The $T$- and $\vec{H}$-dependencies of $\rho$ and $\rho_{yx}$ under were first measured under $\mu_0H\leq\SI{9}{\tesla}$ using a Quantum Design Physical Property Measurement System (PPMS).
Here $H$ denotes the magnitude of the magnetic field $\vec{H}$.

We also measured the transport properties under $\mu_0H\leq\SI{18}{\tesla}$ employing a JASTEC $\SI{18}{\tesla}$ superconducting magnet at High Field Laboratory (HFL), Institute for Material Research, Tohoku University.
We also utilized a $\SI{55}{\tesla}$ pulse magnet at the Center for Advanced High Magnetic Field Science (AHMF), Osaka University to probe the transport properties under extreme $H$.
In these experiments, the measurements of angle-resolved MR with $\vec{H}$ rotating in the $ab$ and the $ac$-plane of the crystals were also conducted with the help of sample rotators.

The magnetoresistance (MR) is defined as:
\begin{align}
  \label{eq:MR}
  \mathrm{MR}(H, \theta, \phi) = \frac{\rho(H, \theta, \phi) - \rho_0}{\rho_0}\,.
\end{align}
Here $\rho_0$ is the resistivity under $H = 0$, and $\theta$ and $\phi$ are defined in Fig.~\ref{fig:intra-theta-phi}(a) and (c), respectively.

We carried out the analyses of the in-plane Hall effect $\rho_{yx}$ as follows. 
At first, we evaluated the carrier numbers using a one-carrier-type model in the $T > \tstar$ regime; 
\begin{align}
  \label{eq:hall_one}
  \rho_{yx} = R_{\mathrm{H}}\mu_{0}H_c\,.
\end{align}
Here $R_{\mathrm{H}}$ is the Hall coefficient and is related to the carrier number $n_{\mathrm{H}}$ and mobility $\mu_{\mathrm{H}}$ as;
\begin{align}
  \label{eq:hall_number}
  n_{\mathrm{H}} = \frac{1}{eR_{\mathrm{H}}}. \\
  \label{eq:hall_mu}
  \mu_{\mathrm{H}} = \frac{1}{en_{\mathrm{H}}\rho_0}.
\end{align} 
The analyses were carried out by fitting Eq.~\eqref{eq:hall_one} to the $\rho_{yx}(H_c)$ experimental data [Fig.~\ref{fig:hall}].
At $T \lesssim \SI{30}{\kelvin}$, where $\rho_{yx}(H_c)$ is strongly non-linear, we employed a linear fit in a small $H_c$ window around $0$, $\mu_0|H_c| < \SI{1}{\tesla}$.

The a two-carrier-type model that was used to analyzed the non-linear $\rho_{yx}(H_c)$ at $T \lesssim \SI{40}{\kelvin}$ is;
\begin{align}
  \label{eq:twocarrier}
  \rho_{yx} = \frac{1}{e}\frac{\mu_h^2\mu_e^2(n_h - n_e)(\mu_0H_c)^3 + (\mu_h^2n_h - \mu_e^2n_e)(\mu_0H_c)}{\mu_h^2\mu_e^2(n_h - n_e)^2(\mu_0H_c)^2 + (\mu_hn_h + \mu_en_e)^2}.
\end{align}
Here,  $n_i$ and $\mu_i$ denote the carrier density and the mobility; the subscripts $e$ and $h$ denote the electron- and the hole- carriers, respectively.

In our analysis, at first we fit a nonlinear $\rho_{yx}(H_c)$ curve by the following equation;
\begin{align}
  \label{eq:fit}
  \rho_{yx} (H_c) = \frac{a (\mu_0H_c)^3 + b H_c}{c (\mu_0 H_c)^2 + 1 }\,.
\end{align}

We note that there are only three fitting parameters in Eq.~(\ref{eq:fit}).
Since the last term in the denominator of the RHS in Eq.~(\ref{eq:fit}) is fixed to $1$, the degrees of freedom in the determination of the other three parameters in Eq.~(\ref{eq:fit}) by fitting are greatly reduced.
Now, using Eq.~(\ref{eq:twocarrier}), $a$, $b$, and $c$ are related to $n_i$'s and $\mu_i$'s as follows.
\begin{subequations}
   \label{eq:params}
  \begin{align}   
    a & = \frac{e \mu_h^2\mu_e^2(n_h - n_e)}{ \sigma_0^2 } \,,\label{eq:param_a}\\
    b & = \frac{e (\mu_h^2n_h - \mu_e^2n_e)}{ \sigma_0^2 } \,,\label{eq:param_b}\\
    c & = \frac{e^2 \mu_h^2\mu_e^2(n_h - n_e)^2 } {\sigma_0^2  } \,,\label{eq:param_c}\\
    \sigma_0 & =  e(\mu_hn_n + \mu_en_e )  \,.  \label{eq:param_sigma0}     
  \end{align}
\end{subequations}
As explained above, $\sigma_0$ in Eq.~(\ref{eq:param_sigma0}) can be measured independently.
By solving the system of equations (\ref{eq:params}), we can obtained $n_i$'s and $\mu_i$'s.

This procedure, fixing one of the fitting parameters, is essential in practice.
  Otherwise, if four parameters had been used in the fitting function in the RHS of Eq.~(\ref{eq:fit}), i.e. the last term in the denominator had been an additional free parameter $d$, they would have not been determined uniquely.
  This is because when the all four parameters are multiplied by a factor, they will produce the same fitting curve, but the obtained values of $n_i$'s and $\mu_i$'s can be spuriously wrong due to relations shown in Eqs~(\ref{eq:params}).

The isobaric specific heat $C_p$ of \bmb was measured in the $T$ range from $\SI{2}{\kelvin}$ to $\SI{200}{\kelvin}$ and under various $B$-strengths using the commercial heat capacity option of PPMS
A single crystalline sample was fixed at the center of the PPMS calorimeter puck by Apiezon N grease so that the direction of $\vec{H}$ was parallel to the $\hat{a}$ axis.
Puck calibration as well as addenda measurements were done at each $B$-strength to minimize the possible errors
The temperature dependence of the specific heat was fitted using the following model;
\begin{align}
  \label{eq:Cp}
  C_p \approx C_v^{\mathrm{D}} + C_v^{\mathrm{E}}\,;
\end{align}
with %
$C_v^{\mathrm{D}} = 9Nk_B\left(\frac{k_BT}{\hbar \omega_{\mathrm{D}}}\right)^3\int_0^{x_{\mathrm{D}}}\frac{x^4\mathrm{e}^x}{(\mathrm{e}^x-1)^2}dx\,$ and %
$C_v^{\mathrm{E}} =  3Nk_B \left(\frac{k_BT}{2\hbar \omega_{\mathrm{E}}}\right)^2 \bigg/ \sinh^2 \left(\frac{k_BT}{2\hbar \omega_{\mathrm{E}}}\right)\,$, %
respectively.
Here $N$ is the number of atoms per unit cell and $k_B$ the Boltzmann constant.
The Debye $\omega_{\mathrm{D}}$ and the Einstein frequency $\omega_{\mathrm{E}}$ were used as fitting parameters and  $x_{\mathrm{D}}$ is the dimensionless Debye number defined as $x_{\mathrm{D}} \equiv \frac{\hbar \omega_{\mathrm{D}}}{k_{\mathrm{B}} T}$. 
The calculation using the model with least-squared algorithm (the Python package Scipy \cite{scipy}) well fitted the experimental data as shown in Fig.~\ref{fig:thermodynamics}(c).

The magnetizations of \bmb \,  single crystals were measured using a Quantum Design Magnetic Property Measurement System (MPMS) along the $ab$-plane and the $\hat{c}$-axis.
In this measurements, samples fixed by GE-vanish onto a holder made from a thin rod of fused silica.

In the framework of the mean field theory, the Hamiltonian of one spin ${\vec{S}}_0$ under the external magnetic field $H_{ext}$ perpendicular to the $\hat{c}$-axis can be described by the exchange interaction and Zeeman energy i.e.
\begin{align}
  \begin{split}
    \mathcal{H}_{spin} = &   (z_1J_1{\vec{S}}_1 + z_2J_2{\vec{S}}_2 + z_c J_c{\vec{S}}_c   )\cdot {\vec{S}}_0  \\
   &- DS_{0,\parallel}^2 + g\mu_B {\vec{S}}_0 \cdot {\vec{H}}_{ext}    
  \end{split}  
  \label{eq:spin-hamiltonian}
\end{align}
Here, ${\vec{S}}_1$, ${\vec{S}}_2$ and ${\vec{S}}_c$ are nearest,second nearest and inter-plane nearest spins. And $J$s are each exchange interaction. the other square and linear terms of ${\vec{S}}_0$ are anisotropic term and Zeeman term.
The second term including ${\bf{S}}_2\cdot{\bf{S}}_0$ make no energy change depending on external field because the angle between these two spins might be zero for any strength of magnetic field. Assuming $\theta$, the angle of spin from $\hat{c}$-axis, the spin can be decomposed to two parts. One is parallel to $\hat{c}$-axis and described by $S\cos\theta$, where $S$ is the magnitude of spin moment. The other is $S\sin\theta$ that is parallel to $H_{ext}$. Due to the antiferromagnetic ordering, the former changes the sign for each nearest neighbor. The Hamiltonian can be rewritten based on these assumptions as
\begin{align}
  \begin{split}
    \mathcal{H}_{spin} = & (z_1J_1+z_cJ_c)S^2(\sin^2\theta-\cos^2\theta) - \\
    & DS^2\cos^2\theta -g\mu_BS\sin\theta H_{ex}
  \end{split}\\
  \begin{split}
    \frac{\partial \mathcal{H}_{spin}}{\partial \theta}
    = & [4(z_1J_1+z_cJ_c)+2D]S^2\sin\theta\cos\theta -
    \\
    & g\mu_BSH_{ex}\cos\theta    
  \end{split}
\end{align}
There is the angle $\theta_0$ that minimize the energy as $\left.\frac{\partial \mathcal{H}_{spin}}{\partial \theta}\right|_{\theta_0} = 0$. As a result,
\begin{align}
  \theta_0 &= \frac{g\mu_B H_{ex}}{4(z_1SJ_1+z_cSJ_c)+2SD} \\
  m_\perp &= g\mu_B S_\perp \approx \frac{g^2\mu_B^2 H_{ex}}{4(z_1SJ_1+z_cSJ_c)+2SD} \\
  \chi_\perp &= \frac{2m_\perp}{H_{ex}} = \frac{2g^2\mu_B }{4(z_1SJ_1+z_cSJ_c)+2SD}[\mu_B/\mathrm{T}/unit\_cell]
\end{align}
Here, $m_\perp$ is a magnetization for one Mn-site and it is doubled when $\chi_\perp$ calculation because one unit cell contain two Mn's. The experimental values $SJ_1 = \SI{21.7}{\milli \electronvolt}, SJ_c = \SI{1.26}{\milli \electronvolt}, SD = \SI{0.046}{\milli \electronvolt}$\cite{calder-2014} result $\chi_{\perp} = 12.96 \times 10^{-3}[\mu_B/\mathrm{T \cdot unit\_cell}]$ that is twice as the experimental result shown in Fig.\ref{fig:thermodynamics}(a). The effective field from exchange interaction also calculated from Eq.~\eqref{eq:spin-hamiltonian} that approximately results $\SI{250}{\tesla}$ as mentioned in Section~\ref{sec:magnetism}.


\section*{Acknowledgements}

  We thank K. Ogushi, T. Aoyama, K. Igarashi, H. Watanabe, Y. Yanase, and T. Arima for fruitful discussions.
  This work was supported by a Grant-in-Aid for Scientific Research on Innovative Areas ``J-Physics'' (Grant No.18H04304), and by JSPS KAKENHI (Grants No. 18K13489, No. 18H03883, No. 17H045326, and No. 18H03858).
  T.O. thanks the financial supports from the International Joint Graduate Program in Materials Science (GP-MS) of Tohoku University.
  Experiments under high magnetic fields was done at High Field Laboratory for Superconducting Materials (HFLSM) in Tohoku University.
  Pulsed field measurements were carried out at the Center for Advanced High Magnetic Field Science in Osaka University under the Visiting Researcher's Program of the Institute for Solid State Physics, the University of Tokyo.
  This research was partly made under the financial support by the bilateral country research program of JSPS between AIMR, Tohoku University and Jozef Stefane Institute, Slovenia.
  This work was also supported by World Premier International Research Center Initiative (WPI), MEXT, Japan.
  DA acknowledges the financial support of the Slovenian Research Agency through BI-JP/17-19-004 and J1-9145 grants.



%
  

\end{document}